\documentclass[aps,pra,twocolumn,amsmath,amssymb,showpacs]{revtex4}
\usepackage{graphicx}
\usepackage{amsmath}
\usepackage{dcolumn}
\usepackage{bm}

\newcommand{\bq}{\mathbf{q}}
\newcommand{\bk}{\mathbf{k}}
\newcommand{\br}{\mathbf{r}}

\newcommand{\aba}{\hat{\mathbf{a}}}
\newcommand{\abb}{\hat{\mathbf{b}}}

\newcommand{\abe}{\hat{\mathbf{e}}}
\newcommand{\abf}{\hat{\mathbf{f}}}

\newcommand{\abn}{\hat{\mathbf{n}}}

\newcommand{\abx}{\hat{\mathbf{x}}}
\newcommand{\aby}{\hat{\mathbf{y}}}
\newcommand{\abz}{\hat{\mathbf{z}}}
\newcommand{\abp}{\hat{\mathbf{p}}}

\newcommand{\abk}{\hat{\mathbf{k}}}
\newcommand{\abr}{\hat{\mathbf{r}}}

\newcommand{\ha}{\hat{ a} }

\begin{document}
\title{Intrinsic Entanglement Degradation by Multi-Mode Detection}
\author{A. Aiello}
\author{J.P. Woerdman}
\affiliation{Huygens Laboratory, Leiden University\\
P.O.\ Box 9504, 2300 RA Leiden, The Netherlands}
\begin{abstract}
Relations between photon scattering, entanglement and multi-mode
detection are investigated. We first establish a general framework
in which one- and two-photon elastic scattering processes can be
discussed, then we focus on the study of the intrinsic
entanglement degradation caused by  a multi-mode detection. We
show that any multi-mode scattered state cannot maximally violate
the Bell-CHSH inequality because of the momentum spread. The
results presented here have general validity and can be applied to
both deterministic and random scattering processes.
\end{abstract}
 \maketitle
%
%
%
%
%

\section{Introduction}
During the last fifteen years, the concept of entanglement has
passed from the inhospitable realm of the ``problems in the
foundation of quantum mechanics'' \cite{Ballentine}, to the
 fashionable fields of quantum information and
computation \cite{NielsenBook}. Experimental availability
\cite{Zeilinger99} of reliable sources of entangled quantum bits
({\em qubits}) in the optical domain has been one of the major
boosts for this rapid transition.
Polarization-entangled photon pairs produced in a spontaneous
parametric down conversion (SPDC) process \cite{YarivBook} have
thus become an essential tool in experimental quantum information
science \cite{Gisin02}.

Robustness of photon polarization-entanglement is important when
propagation over long distances of entangled states is required.
However, during propagation in a medium  photons may be
depolarized by {\em scattering} processes due to imperfections.
Further, the recently observed transfer of quantum entanglement
from photons to a linear medium and back to photons
\cite{Altewischer02}, has also been described as an elastic
scattering process \cite{Scatheory}.
In general, polarization-dependent scattering may affect more or
less the entanglement of a photon pair depending on the properties
of the scattering medium \cite{Velsen04,Woerdman03}.

In this paper we show that independently from the details of the
scattering process, the polarization entanglement of a photon pair
is unavoidably degraded by a multi-mode measurement
\cite{Woerdman03}. Moreover, we show that, contrary to  common
belief, it is not always possible to build a $2 \times 2$ reduced
density matrix (or $4 \times 4$ in the case of entangled pairs) to
describe the scattered state. This fact was recently recognized by
Peres and coworkers \cite{Peres_et_al} for the case of a {\em
single} polarization qubit, as opposed to the case of two-photon
{\em entangled} qubits considered here. They circumvented this
difficulty by introducing, at a certain stage of their
calculations, the unphysical concept of longitudinal photons. In
this way they were able to build a $3 \times 3$ effective reduced
density matrix whose physical content can be obtained from a
family of positive operator-valued measures (POVMs
\cite{PeresBook}).

In this paper we follow a somewhat different approach. We show
that the difficulties one encounters in trying to define a reduced
density matrix, are also due to the troubling determination of the
effective dimension of the Hilbert space in which the two-photon
entangled state can be represented. In fact, due to the transverse
nature of the free electromagnetic field, in a multi-mode
scattering process the momentum and the polarization degrees of
freedom become entangled. Therefore, after the scattering, the
two-photon polarization-entangled state is no longer confined to a
$4$-dimensional Hilbert space, but spans an higher-dimensionality
space. We show that under certain circumstances this dimension may
remain bigger than $4$ even after tracing out the momentum degrees
of freedom. In this case a traditional Bell-violation measurement
setup may fail to reveal a maximal violation of the Bell-CHSH
inequality. However, the detailed investigation of the
entanglement properties of this kind of high-dimensional scattered
photon-pair \cite{Collins02,Law00,Law04}, is left to a forthcoming
paper \cite{Aiello04_2}.

This paper is structured as follows.
 In Sec.
II we define the general framework for the discussion of one- and
two-photon scattering processes.
 In Sec.
III,  starting from an experimentalist point of view, we adopt an
operational way to define the photon polarization states in terms
of the physically detected states. Furthermore, we calculate the
classical Jones matrix \cite{KligerBook} corresponding to an
arbitrarily oriented polarizer and use it to introduce a suitable
field creation operator to build explicitly the physical photon
states. In this way we can introduce a
 $2 \times 2$ effective density matrix by a projection
operation on the one-photon physical states.
The concept of effective density matrix is extended to
polarization-entangled photon-pair states, in the first part of
Sec. IV. Then, in the second part of  Sec. IV we quantify the
entanglement degradation by calculating the effect of a
polarization detection mismatch on the mean value $\langle
\mathcal{B}_{\mathrm{CHSH}} \rangle$ of the Bell-CHSH operator
\cite{Braunstein92} for the set of  the Schmidt states
\cite{Aravind95}. Finally, we summarize our main conclusions in
Sec. V.
\section{One-Photon Scattering}
\subsection{One-photon states and their representation}
To understand the effects that a scattering process may have on a
single photon wave packet, we start by introducing a proper
notation.
Let ${\mathbf{A}}(\br, t)$ denote the transverse part of the
electromagnetic potential. In the Coulomb gauge the quantization
procedure can be carried out straightforwardly \cite{LeeBook} to
obtain the corresponding field operators $\hat{\mathcal{A}}(\br,
t)$.
At any given time $t$, $\hat{\mathcal{A}}(\br, t)$ may be expanded
in terms of the Fourier series
\begin{equation}\label{2.3}
\hat{\mathcal{A}}(\br, t) = \sum_{\bk} \frac{1}{\sqrt{2 \omega
\Omega}} \left[ \aba_{\bk}(t) e^{i \bk \cdot \br}  +\mathrm{ h.c.}
\right],
\end{equation}
where $\omega = |\bk|$, $\aba_\bk(t)\cdot \bk = 0$, and natural
units $(c = \hbar =1)$ have been used. With $\Omega$ we denote the
quantization volume. For any given $\bk$ we may define a set of
three real orthogonal unit vectors $\abk \equiv \bk/|\bk|$,
$\abe_\bk^{(1)}$, $\abe_\bk^{(2)}$ such that
\begin{equation}\label{3.3}
\abe_\bk^{(1)} \cdot \abe_\bk^{(2)}=0, \qquad \abe_\bk^{(1)}
\times \abe_\bk^{(2)} = \abk.
\end{equation}
It is convenient to introduce the creation operators
$\ha^\dagger_{\bk s}$ of a photon with momentum $\bk$ and helicity
$s=\pm 1$ as
\begin{equation}\label{1.4}
\ha^\dagger_{\bk s} =  \aba^\dagger_{\bk} \cdot \abf_\bk^{(s)},
\end{equation}
where the complex vectors $ \abf_\bk^{(s)}$ are defined as
\begin{equation}\label{1.4.1}
\abf_\bk^{(s)} = \frac{1}{\sqrt{2}} \left( \abe_\bk^{(1)} +  i s
\, \abe_\bk^{(2)}\right), \qquad (s = \pm1).
\end{equation}
 They satisfy the canonical commutation rules
\begin{equation}\label{2.4}
\left[\ha_{\bk s}(t), \ha^\dagger_{\bk' s'} (t) \right] =
\delta_{\bk \bk' } \delta_{s s'}.
\end{equation}
Then, an one-photon helicity basis state can be written as
\begin{equation}\label{3.4}
| \bk, s(\bk) \rangle \equiv \ha^\dagger_{\bk s} | 0 \rangle,
\end{equation}
where $| 0 \rangle $ denotes the vacuum state of the {\em free}
electromagnetic field.
%
This state represents a photon with  momentum $\bk$ and helicity
$s(\bk) =\pm1$, that is $| \bk, s(\bk) \rangle$ is an eigenstate
of the momentum operator $\hat{\mathbf{K}}\equiv \sum_{\bk',s'}
\bk' \ha^\dagger_{\bk' s'} \ha_{\bk' s'} $ with eigenvalue $\bk$:
\begin{equation}\label{3.5}
\hat{\mathbf{K}} | \bk, s(\bk) \rangle = \bk | \bk, s(\bk)
\rangle.
\end{equation}
Because of Eq. (\ref{3.5}), the one-photon-space basis states $ |
\bk, s(\bk) \rangle$ are often said to be the direct products of
momentum and polarization states and this is usually stated by
writing
\begin{equation}\label{b2}
|\bk, s(\bk)\rangle \equiv | \bk \rangle \otimes | s(\bk) \rangle.
\end{equation}
However, one should realize that the left side of this equation is
a genuine QED state, while the right side is just one of its
possible representations. To be more specifics in QED there is no
such a thing  like a momentum creation operator
that creates from the vacuum a  momentum state $|\bk\rangle$, nor
is there a polarization creation operator
which create a polarization state $|s(\bk) \rangle$.
Nevertheless the ``momentum $\otimes$ polarization''
representation for the one-photon states has several advantages.
For this reason we now  present  formally a few elementary facts
about this representation.
\subsubsection{Some elementary facts}
Let us consider  two one-photon states $|\psi_\bk\rangle$ and
$|\phi_\bq\rangle$  which are eigenstates of the momentum operator
with eigenvalues $\bk$ and $\bq$ respectively:
\begin{equation}\label{b4}
\begin{array}{rcl}
|\psi_\bk\rangle &= &\psi_+ (\bk) |\bk, 1\rangle +  \psi_- (\bk)
|\bk, -1\rangle, \\\\
|\phi_\bq\rangle &=& \phi_+ (\bq) |\bq, 1\rangle +  \phi_- (\bq)
|\bq, -1\rangle.
\end{array}
\end{equation}
They can be represented as
\begin{equation}\label{b5}
|\psi_\bk \rangle \doteq \left(
\begin{array}{c}
  \psi_+ \\
  \psi_-
\end{array}
 \right)_\bk, \qquad |\phi_\bq\rangle \doteq \left(
\begin{array}{c}
  \phi_+ \\
  \phi_-
\end{array}
 \right)_\bq,
\end{equation}
where the symbol ``$\doteq$'' stands for ``is represented by'',
and we have used the subscripts $\bk$ and $\bq$ for the
representations of $|\psi_\bk\rangle$ and  $|\phi_\bq\rangle$
respectively, to stress the fact that they belong to different
vector spaces. Since
\begin{equation}\label{b6}
\left(
\begin{array}{c}
  \psi_+ \\
  \psi_-
\end{array}
 \right)_\bk = \psi_+ \left(
\begin{array}{c}
  1 \\
 0
\end{array}
 \right)_\bk +  \psi_- \left(
\begin{array}{c}
  0 \\
 1
\end{array}
 \right)_\bk,
\end{equation}
we can write
\begin{equation}\label{b7}
|\psi_\bk \rangle \doteq \psi_+ | + )_\bk + \psi_-|- )_\bk,
\end{equation}
where we have defined
\begin{equation}\label{b8}
| + )_\bk \doteq \left(
\begin{array}{c}
  1 \\
 0
\end{array}
 \right)_\bk, \qquad  |- )_\bk \doteq \left(
\begin{array}{c}
  0 \\
 1
\end{array}
 \right)_\bk.
\end{equation}
In the equations above we have used the parenthesis symbols $ |
\pm )_\bk$ for the basis vectors instead of the usual bracket $|
\cdot \rangle$ ones  to underline the fact that they are not
genuine QED states, that is they are not created from the vacuum
by any bosonic creation operator. Then, from now on, we shall
reserve the bracket symbols $| \cdot \rangle$  only for the truly
QED states.

The two vectors displayed in Eq. (\ref{b8}) form a basis
$\mathcal{B}_\bk$ for
 a two-dimensional ``polarization'' space $\mathcal{H}_\bk$.
 Exactly in the same way we can introduce a basis $\mathcal{B}_\bq = \{   |+)_\bq,  |-
)_\bq\}$ for the space $\mathcal{H}_\bq$ relative to the
(arbitrary) momentum $\bq$. More generally, let $K$ be  a
$N$-dimensional set of momenta: $K = \{\bk_1, \ldots, \bk_N
 \}$. Then we can build a $2N$-dimensional space
 $\mathcal{H}$ spanned by the basis $\mathcal{B} = \mathcal{B}_{\bk_1} \cup
 \dots\cup
 \mathcal{B}_{\bk_N}$
as
\begin{equation}\label{b9}
  \mathcal{H} = \mathcal{H}_{\bk_1} \oplus \dots \oplus
  \mathcal{H}_{\bk_N}.
\end{equation}
It is clearly possible to introduce other bases in $\mathcal{H}$.
For example let $\{ |\bk_i) \}$, ($i = 1,\ldots,N$) and $\{|+ ),
|- ) \}$ be a $N$-dimensional and a $2$-dimensional sets of basis
vectors respectively, represented by
\begin{equation}\label{b10}
|\bk_i ) \doteq \left( \begin{array}{c}
  0 \\
  \vdots \\
 1\\
 \vdots\\
 0
\end{array} \right), \quad | + )\doteq \left(
\begin{array}{c}
  1 \\
 0
\end{array}
 \right), \quad  |- )\doteq \left(
\begin{array}{c}
  0 \\
 1
\end{array}
 \right).
\end{equation}
where all elements of $|\bk_i ) $ are zero with the exception of
element $i$ which is equal to $1$. Then the set given by the
vectors
\begin{equation}\label{b11}
|\bk_i ) \otimes |s ) \equiv  |\bk_i, s ), \qquad
\left(\begin{array}{c}
 i = 1, \ldots, N,\\s = \pm 1,
\end{array}\right)
\end{equation}
is an orthonormal and complete basis of $\mathcal{H}$. Now, the
factorization ``momentum $\otimes$ polarization'' in Eq.
(\ref{b11}) is absolutely legitimate without introducing any ``ad
hoc'' creation operator. We would like to stress once again the
fact that the genuine QED states $| \bk, s(\bk) \rangle$ {\em are
not} factored, just their representation $| \bk, s )$. Moreover,
the helicity label $s$  in Eq. (\ref{b11}) does not  depend on the
momentum $\bk$, so it is not longer a secondary variable
\cite{Peres_et_al}. Note that we have chosen to work with the
helicity eigenstates $| \bk, s(\bk) \rangle$ but we could have
worked equally well with the linear polarization basis vectors  $|
\bk, \alpha(\bk) \rangle$, where $\alpha  = x,y$.
\subsubsection{Tracing over the momenta}
The reason for which we have introduced the factorizable basis
vectors $|\bk, s )$ in the previous subsection, is that we want to
be able to separate the momentum degrees of freedom from the
polarization ones. This separation becomes important when the
photons are subjected to  polarization tests only, and the momenta
become irrelevant degrees of freedom. Then, it is a standard
practice \cite{PeresBook} to introduce a reduced density matrix
obtained from the complete one, by tracing over some subset
 of the whole set of the field momenta.

Let $\hat{\rho}$ be the density matrix describing the state of a
single photon.
 We can
represent $\hat{\rho}$ in $\mathcal{H}$ by defining the $N^2$
matrices $R(n,m)$ ($2 \times 2$) whose elements are calculated
with the usual QED rules:
\begin{equation}\label{b12}
{R}_{s s'}(n,m)\equiv Z_1 \langle \bk_n, s(\bk_n)| \hat{\rho} |
\bk_m, s'(\bk_m) \rangle,
\end{equation}
where $Z_1$ is an arbitrary normalization constant and $(s,s' =
\pm 1 )$. These matrices are the building blocks of the total
representation of $\hat{\rho}$:
\begin{equation}\label{b13}
\hat{\rho} \doteq  \left(
\begin{array}{ccc}
  \left[ \begin{array}{cc}
    R_{++}(1,1) & R_{+-}(1,1) \\
    R_{-+}(1,1) & R_{--}(1,1) \
  \end{array}\right] & \dots &  R(1,N) \\
  \vdots &\ddots & \vdots \\
   R(N,1) & \dots &  R(N,N)
\end{array}
\right).
\end{equation}
From now on, we choose $Z_1$ such that $\mathrm{Tr}\hat{\rho}=1$.
Now, if we want to calculate the {\em reduced} density matrix
$\hat{\rho}^\mathrm{R}_1$, obtained by tracing out the momentum
degrees of freedom, what we have to do is simply to take all the
diagonal blocks $R(n,n)$ and sum them together:
\begin{equation}\label{b14}
\hat{\rho}^\mathrm{R}_1 \doteq \sum_{n=1}^N R(n,n) \equiv
\rho^\mathrm{R}_1.
\end{equation}
The matrix $\rho^\mathrm{R}_1$ is a well defined $2 \times 2$
matrix and there is no ambiguity in its determination. We have
used the subscript $1$ in writing $\rho^\mathrm{R}_1$ to
distinguish it from the generalized reduced density matrix
$\rho^\mathrm{R}$  we shall introduce in the next subsection.
However, it was recently shown that a reduced density matrix
introduced as above, has no meaning because it does not have
definite transformation properties \cite{Lindner04}. Besides this
fact, there are more practical reasons for which
$\rho^\mathrm{R}_1$ (at least a single one) cannot be introduced,
as we shall see in the next subsection. For the moment let us see
very shortly how the traditional approach works and why it may
fail. If one deals with operators that are factorizable in the
basis $\{|\bk_i, s)\}$, that is if they can be represented as
\begin{equation}\label{new1}
\hat{X} \doteq |X)(X| \otimes \sum_{n=1}^N|\bk_n )(\bk_n|,
\end{equation}
where the right side of this equation is, in fact, the direct
product of a $2 \times 2$ matrix (polarization part, momentum
independent), times a $N \times N$ matrix (momentum part,
polarization independent). Then the average value $\langle \hat{X}
\rangle = \mathrm{Tr}(\hat{\rho} \hat{X})$ can be certainly
written as
\begin{equation}\label{new2}
\bigl\langle \hat{X} \bigr\rangle =
\mathrm{Tr}\bigl(\hat{\rho}^\mathrm{R}_1 |X)(X| \bigr),
\end{equation}
and, apparently, there are no problems in using
$\hat{\rho}^\mathrm{R}_1$. The key point here, is that the
operator representation  with respect to the factorizable basis
$\{| \bk_n, s) \}$ may have no meaning at all because the
polarization state $| X )$ in Eq. (\ref{new1}) does not depend on
$\bk$ while the polarization state of a photon is {\em always}
referred to a given momentum, as shown in Eq. (\ref{1.4}). Then,
in a multi-mode process where many values of the photon momentum
are involved, it  becomes impossible to define a unique $2 \times
2$ reduced density matrix ($ |X)(X|$ in Eq. (\ref{new1})), instead
one must deal with a different matrix for each different momentum.
This will be explicitly shown in the next subsection.
\subsection{One-Photon Scattering}
Let us consider now a one-photon state approximatively represented
by the monochromatic plane wave $| \bk_0, s_0(\bk_0)
\rangle_{\mathrm{in}}$, and suppose that it is elastically
scattered by a certain medium. The final state $| \psi_f \rangle$
of the photon after the scattering process can be written in terms
of the output states $| \bk, s(\bk) \rangle$ as
\begin{equation}\label{5.4}
| \psi_f \rangle = \sum_{\bk \in K}  \sum_{s = \pm1}
\psi_{s}(\bk)| \bk, s(\bk) \rangle,
\end{equation}
where $\psi_{s}(\bk) \equiv \langle \bk, s(\bk) | \bk_0,
s_0(\bk_0) \rangle_{\mathrm{in}}$ represent the probability
amplitude that the photon is scattered in a state with momentum
$\bk$ and helicity $s(\bk)$, where  $ |\bk| \cong \omega_0$. With
$K$ we have denoted the set of all scattered modes. By inspecting
Eq. (\ref{5.4}) one  immediately realizes that in the scattered
state $|\psi_f \rangle$ the polarization and momentum degrees of
freedom are entangled because the helicity states $|s(\bk)\rangle$
depend explicitly from $\bk$ and the two sums are not independent.
In other words, in general, $|\psi_f \rangle$ is not an eigenstate
of the linear momentum operator $\hat{\mathbf{K}}$.

Now suppose that we  measure some polarization property of the
scattered photon, regardless of its momentum. To be more specific
let us assume that a polarization analyzer (a dichroic sheet
polarizer or a crystal prism polarizer) is present in a plane
perpendicular to the wave-vector $\bk_0$ of the impinging photon
and that we collect, with a photo-detector,  all the light coming
from the scattering target within a certain angular aperture
$\Theta_D$. With this experimental configuration we test only the
photon polarization, irrespective of the plane-wave mode $\bk$
where the photon is to be found within $\Theta_D$. We call this
kind of experimental arrangement, a {\em multi-mode} detection
scheme.

The act of measurement can be quantified by calculating, for
example, the mean value of the operator $\hat{P}$ defined as:
\begin{equation}\label{b30}
 \hat{P} = \sum_{\bk \in K_D} \sum_{s, s'} P_{s s'}(\bk) | \bk,
 s(\bk) \rangle \langle \bk, s'(\bk) |,
\end{equation}
where
\begin{equation}\label{b40}
P(\bk) = \left( \begin{array}{cc}
  P_{++}(\bk) &  P_{+-}(\bk) \\
   P_{-+}(\bk) &  P_{--}(\bk)
\end{array} \right),
\end{equation}
is a $2 \times 2$ Hermitian matrix such that $P^2 = P$ and $K_D$
is the set of the detected modes:
 \[ \{ \bk \in K_D | \, \left| \bk \cdot \bk_0
\right|/\omega_0^2 \lesssim |\cos \Theta_D |\}.\]
 Moreover we
assume, for definiteness, $\dim K_D = N$ and $K_D \subset K$.
It is easy to see that $\hat{P}$ is a projector ($\hat{P}^2 =
\hat{P}$), and that it commutes with the momentum operator
$\hat{\mathbf{K}}$:
\begin{equation}\label{new10}
\bigl[\hat{P}, \hat{\mathbf{K}} \bigr] = \mathbf{0}.
\end{equation}
It is also easy  to understand that the equation above is the
equivalent, in the QED context, of the factorizability condition
expressed in Eq. (\ref{new1}).
The matrix elements of $\hat{P}$ are
\begin{equation}\label{b50}
\begin{array}{rcl}
P_{s s'}(n,m) & \equiv & \langle \bk_n, s(\bk_n)| \hat{P} | \bk_m,
s'(\bk_m) \rangle \\\\
 & \equiv & P_{s s'}(n) \delta_{nm}.
\end{array}
\end{equation}
From Eq. (\ref{b50}) is clear that $\hat{P}$ has a block-diagonal
shape:
\begin{equation}\label{b60}
\hat{P} \doteq \left(
\begin{array}{ccc}
  \left[ \begin{array}{cc}
    P_{++}(1) & P_{+-}(1) \\
    P_{-+}(1) & P_{--}(1) \
  \end{array} \right] &  &   \\
  &\ddots &  \\
  &  &   P(N)
\end{array}
\right),
\end{equation}
so that only the corresponding diagonal blocks of $\hat{\rho}$
will enter in the calculation of $\langle \hat{P}\rangle$.
Explicitly
\begin{equation}\label{b70}
\langle \hat{P}\rangle = \sum_{n=1}^N \mathrm{Tr} \bigl(R(n)
P(n)\bigr),
\end{equation}
where Eq. (\ref{b13}) has been used and $R(n) \equiv R(n,n)$ for
short.
Then, the form of the operator $\hat{P}$ in Eq. (\ref{b60})
naturally leads to the introduction of a {\em generalized} reduced
density matrix ${\rho}^\mathrm{R}$ ($2N \times 2N $)
 defined as:
\begin{equation}\label{2.7}
{\rho}^\mathrm{R} = \mathrm{diag}  \{R(1), \ldots, R(N) \} =
\left(
\begin{array}{ccc}
R(1) &  &  \\
   & \ddots &  \\
   &  & R(N)
\end{array}
\right),
\end{equation}
 Thus, differently from Eq. (\ref{b14}) were we have found a
unique $2 \times 2$ matrix, we have now to deal with a set of $N$
different $2 \times 2$ matrices $\{ R(1), \ldots, R(N) \}$: one
matrix $R(n)$ per each mode $\bk_n$ of the field. This
 is due to the form of the operator
$\hat{P}$. Only in the hypothetical case in which the matrix
elements $P_{s s'}(\bk)$ of $\hat{P}$  were independent from
$\bk$, we would obtain again a unique $2 \times 2$ density matrix
defined as $\rho^\mathrm{R}_1 = \sum_{n = 1}^N R(n)$.

 What does all this mean?
We can get some insights into the above equations, by rewriting
them in the basis $\mathcal{B} = \mathcal{B}_{\bk_1} \cup \dots
\cup  \mathcal{B}_{\bk_N}$ introduced previously. Let us start by
defining the normalized state $| \phi)_\bk$ in the basis
$\mathcal{B}_{\bk}$ as
\begin{equation}\label{new30}
| \phi)_\bk = \frac{1}{\sqrt{\zeta(\bk)}} \sum_{s = \pm1}
\psi_s(\bk) | s )_\bk,
\end{equation}
where the complex amplitudes $\psi_s(\bk)$ are defined in Eq.
(\ref{5.4}) and
\begin{equation}\label{new35}
\zeta(\bk) \equiv |\psi_+(\bk)|^2 + |\psi_-(\bk)|^2.
\end{equation}
Then we can represent the one-photon state $|\psi_f\rangle$ in Eq.
(\ref{5.4}), restricted to $K_D$, as
\begin{equation}\label{new40}
  | \psi_f \rangle \doteq \bigcup_{\bk\in K_D} | \phi)_\bk.
\end{equation}
Then, by combining Eq. (\ref{b12}) with Eq. (\ref{5.4}) we find
\begin{equation}\label{new50}
R_{s s'}(n,m) = Z_1 \psi_s(\bk_n) \psi_{s'}^*(\bk_m).
\end{equation}
From the equation above and the normalization condition
$\mathrm{Tr}R =1$, it follows that
\begin{equation}\label{new60}
   Z_1 = 1/\sum_{\bk \in K_D} \zeta(\bk).
\end{equation}
It should be clear now that if we define the pure-state density
matrix $\hat{\rho}_\phi(\bk)$ as
\begin{equation}\label{new70}
\hat{\rho}_\phi(\bk) \equiv  | \phi)_\bk \, \/_\bk \! ( \phi |,
\end{equation}
then we can write the generalized reduced density matrix
$\hat{\rho}^\mathrm{R}$ as the {\em direct sum}, as opposed to the
ordinary sum in Eq. (\ref{b14}), of all the sub-matrices
$\hat{\rho}_\phi(\bk)$:
\begin{equation}\label{2.6}
\begin{array}{rcl}
\hat{\rho}^\mathrm{R} & =& \displaystyle{ {\bigoplus_{\bk \in
K_D}} w(\bk)
\hat{\rho}_\phi(\bk)}\\\\
& =& \displaystyle{ {\bigoplus_{\bk \in K_D}} | \phi)_\bk \,
w(\bk) \, \/_\bk \! ( \phi | },
\end{array}
\end{equation}
were the statistical weight function $w(\bk)$ is equal to
\begin{equation}\label{new80}
w(\bk) = Z_1 \zeta(\bk) =
\frac{\zeta(\bk)}{\displaystyle{\sum_{\bk'\in K_D} \zeta(\bk')}},
\end{equation}
and $\displaystyle{\sum_{\bk\in K_D}}w(\bk) =1$. So we find that
the price to pay for introducing a polarization basis
$\mathcal{B}$, is that we have to deal with a generalized reduced
density matrix which is expressed as a direct sum instead of an
ordinary sum. Of course, this fact is entirely  due to the
incomplete nature of the polarization representation; there are no
problems in writing $\hat{\rho}^\mathrm{R}$ as an ordinary sum in
the QED formalism. In such a context we have to use the complete
momentum eigenstate basis $\{ | \bk, s(\bk) \rangle \}$ in order
to write
\begin{equation}\label{new90}
 \hat{\rho}^\mathrm{R} = \sum_{\bk \in K_D}  w(\bk)  | \bk, \phi(\bk)  \rangle
 \langle \bk, \phi(\bk) |,
\end{equation}
where we have defined the momentum eigenstates $| \bk, \phi(\bk)
\rangle$ as
\begin{equation}\label{new100}
| \bk, \phi(\bk)  \rangle \equiv
\frac{1}{\sqrt{\zeta(\bk)}}\sum_{s=\pm 1}\psi_s(\bk) | \bk,
 s(\bk) \rangle  .
\end{equation}
Not surprisingly, $\hat{\rho}^\mathrm{R}$ has the same
``momentum-diagonal'' structure as $\hat{P}$ in Eq. (\ref{b30}).
It clearly represents a {\em mixed} state because
$(\hat{\rho}^\mathrm{R})^2 \neq \hat{\rho}^\mathrm{R}$. Eq.
(\ref{new90}) can be interpreted by saying that from the observer
point of view (who supposedly cannot measure the photon momentum),
things go as if the scatterer were a thermal source emitting
photons in the states $| \bk, s(\bk) \rangle$ with probabilities
$w(\bk)$.

Thus, we have shown that it is impossible to extract from
$\hat{\rho}^\mathrm{R}$ in a straightforward way a unique $2
\times 2$ reduced density matrix; rather we have obtained a
$N$-dimensional set $\{R(1), \dots, R(N) \}$ of $2 \times 2$
matrices,  each of them well defined in its proper Hilbert space
specified by the momentum $\bk$. This is a direct consequence of
the fact that helicity of a photon can be defined only with
respect to a given momentum \cite{Peres_et_al}. It is worth to
note that this result is not in contrast with existence of a $4
\times 4$ density matrix for two polarization-entangled photons
emitted in a SPDC process. In fact, in that case, each photon is
emitted in {\em one} mode of the electromagnetic field and the
two-photon state is still an eigenstate of the total linear
momentum. However, in Sec. III we shall introduce an {\em
effective} reduced density matrix $\rho_\mathrm{eff}$ both for the
one-photon and two-photon states.

In order to perform an actual calculation, we have to give a rule
how to obtain the matrix elements $P_{s s'}(\bk)$ in Eq.
(\ref{b30}). From a physical point of view they can be expressed
as suitable linear combinations of  the probability amplitude that
a photon with definite momentum $\bk$ and linear polarization
$\alpha(\bk)$, is transmitted across a linear polarizer with a
given orientation $\hat{\mathbf{p}}$. The next Section will be
devoted to the search of an explicit expression for $P_{s
s'}(\bk)$.

\section{Physical Polarization States}
In this section we study how a single-photon state changes when
the photon crosses an arbitrarily oriented polarizer. Then we use
this information to build an effective reduced density matrix.
\subsection{Classical Polarization States}
A lossless linear polarizer \cite{BornWolf} is a planar device
which can be characterized by two orthogonal vectors: the {\em
axis} $\abn$ and the {\em orientation} $\abp$. The first vector is
orthogonal to the plane of the polarizer, while the second one
lies in that plane: $\abp \cdot \abn = 0$. From now on, with the
sentence ``a polarizer $(\abp,\abn)$'' we shall indicate a linear
polarizer with orientation $\abp$ and axis $\abn$.

For the moment we consider only {\em classical} fields, later we
shall introduce the corresponding quantum operators. Following
Mandel and Wolf \cite{MandelBook}, we consider a polarizer $(\abp,
\abz)$ where $\abp = \abx \cos \beta + \aby \sin \beta$ and $\abx,
\aby, \abz$ form a cartesian frame. Let $\mathbf{E}^{I}$ and
$\mathbf{E}^{T}$ denote the incident and the transmitted electric
field, respectively. We assume that $\mathbf{E}^{I}$ and
$\mathbf{E}^{T}$ are plane waves propagating in the direction
$\abz'$. Then we can write explicitly:
\begin{eqnarray}
  \mathbf{E}^I & = & E_{x'}^I \abx' + E_{y'}^I \aby',\label{app1I} \\
  \mathbf{E}^T & = & E_{x'}^T \abx' + E_{y'}^T \aby',\label{app1T}
\end{eqnarray}
where $\abx', \aby', \abz'$ are three orthogonal unit vector. If
$\theta$ and $\phi$ are the spherical coordinate of $\abz'$ with
respect to $\abx, \aby, \abz$ (where $\abz$ is the axis of the
polarizer), then
\begin{eqnarray}
  \abx'  & = & \abx \cos \theta  \cos \phi  + \aby \cos \theta  \sin \phi  - \abz \sin \theta ,\label{app2x} \\
  \aby'  & = &  -\abx  \sin \phi  + \aby \cos \phi , \label{app2y} \\
   \abz' & = & \abx \sin \theta  \cos \phi  + \aby \sin \theta  \sin \phi  + \abz \cos
  \theta\label{app2z},
\end{eqnarray}
and $\theta < \pi/2$.
The action of the polarizer can be found by requiring that the
polarization $\abe^{(p)}$ of the transmitted field, lies in the
plane defined by the polarizer orientation $\abp$ and the
propagation vector $\abz'$ of the impinging field \cite{Fainman}.
In other words, $\abe^{(p)}$ can be written as a linear
combination of $\abp$ and $\abz'$: $\abe^{(p)} = c_1 \abp + c_2
\abz'$. The two real coefficients $c_1$ and $c_2$ can be found by
imposing normalization: $|\abe^{(p)}|=1$, and orthogonality:
$\abe^{(p)} \cdot \abz' = 0$. The final result is
\begin{equation}\label{app3}
D(\beta) \abe^{(p)} = \abp - \abz'  ( \abz' \cdot \abp),
\end{equation}
where $D(\beta) \equiv [1 - ( \abz' \cdot \abp)^2 ]^{1/2}$. For
later purposes, it is useful to write explicitly $\abe^{(p)}$ in
terms of its components:
\begin{equation}\label{app4}
\abe^{(p)} =  \abx' \left( \frac{\abp \cdot \abx'
}{D(\beta)}\right) + \aby' \left( \frac{\abp \cdot \aby'
}{D(\beta)} \right).
\end{equation}
Finally, we require that the  transmitted field equals the
projection along $\abe^{(p)}$ of the impinging one:
\begin{equation}\label{app5}
\mathbf{E}^T = (\mathbf{E}^I \cdot \abe^{(p)}) \abe^{(p)}.
\end{equation}
Eq. (\ref{app5}) defines completely the action of the polarizer on
the field. Now substituting  Eq. (\ref{app1T})  and Eq.
(\ref{app4}) in the left and right side of Eq. (\ref{app5})
respectively, and by using Eq. (\ref{app1I}), we obtain
\begin{equation}\label{app6}
  \mathbf{E}^T =\mathbf{E}^I T,
\end{equation}
where we have represented the impinging and transmitted fields as
the row vectors $ \mathbf{E}^I$ and $ \mathbf{E}^T$ respectively,
and the transmission  Jones \cite{KligerBook} matrix  $T$ of the
polarizer is
\begin{equation}\label{app7}
\begin{array}{lcl}
T & = & \displaystyle{ \left( \begin{array}{cc}
     \frac{(\abx' \cdot \abp)^2}{D^2} & \frac{(\abx' \cdot \abp)(\aby' \cdot \abp)}{D^2}
     \\\\
     \frac{(\abx' \cdot \abp)(\aby' \cdot \abp)}{D^2} & \frac{(\aby' \cdot \abp)^2}{D^2} \\
   \end{array} \right),}\\\\
   & = &
\displaystyle{ \left(
\begin{array}{cc}
   \frac{ \cos^2  \theta  \cos^2 \bar{\beta} }{1 - \sin^2  \theta  \cos^2 \bar{\beta} } & \frac{ \cos  \theta  \sin \bar{\beta} \cos \bar{\beta} }{1 - \sin^2  \theta  \cos^2 \bar{\beta} }\\\\
      \frac{ \cos  \theta \sin \bar{\beta} \cos \bar{\beta} }{1 - \sin^2  \theta  \cos^2 \bar{\beta} } & \frac{ \sin^2 \bar{\beta} }{1 - \sin^2  \theta
 \cos^2 \bar{\beta} }\\
   \end{array} \right),}
   \end{array}
\end{equation}
where $\bar{\beta} \equiv \beta -\phi$.
 It is
easy to check that $T$ is a real symmetric projection operator,
and therefore $T T = T$. Moreover for $\theta = 0 =\phi$, Eq.
(\ref{app7}) reduces to the standard literature result for an
''orthogonal'' polarizer $(\abp,\abz')$ (see, e.g., ref.
\cite{MandelBook}, Eq. (6.4-20)).
\subsection{Quantum Polarization States}
Now we  translate  this in  quantum language. Since we have
introduced the generalized Jones matrix $T$ for a {\em linear}
polarizer, it is convenient to define the creation operators
$\hat{a}^\dagger_{\bk \alpha}$ of a photon with momentum $\bk$ and
linear polarization $\alpha = 1,2$ as
\begin{equation}\label{linear}
\ha^\dagger_{\bk \alpha} = \aba^\dagger_{\bk} \cdot
\abe_\bk^{(\alpha)}, \qquad ( \alpha = 1,2) .
\end{equation}
Since $\bk / |\bk| = \abz'$ and the photon momentum is not
affected by the polarizer, we define, for the sake of simplicity,
$\ha_{x'}\equiv \ha_{\bk 1}$, $\ha_{y'}\equiv \ha_{\bk 2}$.
Therefore, from now on we shall always omit the momentum
dependence in the expression of the one-photon states, only the
bracket symbol $| \cdot \rangle$ will remind that we are dealing
with truly QED states.

Again, we follow Mandel and Wolf \cite{MandelBook} and introduce
the operator vector field amplitude $\hat{\mathbf{a}}^I$ such that
\begin{equation}\label{app8}
\hat{\mathbf{a}}^I = \hat{a}_{x'} \abx' + \hat{a}_{y'} \aby'.
\end{equation}
The vector field amplitude $\hat{\mathbf{a}}^T$ behind the
polarizer, is determined by using the transformation law
\begin{equation}\label{app9}
\hat{\mathbf{a}}^T = \hat{a}_{i'} T_{i' x'} \abx' + \hat{a}_{i'}
T_{i' y'} \aby', \qquad (i = x,y),
\end{equation}
where summation over repeated indices is understood. In explicit
form Eq. (\ref{app9}) reads:
\begin{equation}\label{app10}
  \hat{\mathbf{a}}^T =  \abe^{(p)} \hat{b}_{p},
\end{equation}
where we have defined:
\begin{equation}\label{app11}
\hat{b}_p \equiv  \hat{a}_{x'} \left( \frac{\abp \cdot \abx'
}{D(\beta)} \right) + \hat{a}_{y'} \left( \frac{\abp \cdot \aby'
}{D(\beta)} \right).
\end{equation}
Reminding that $D^2(\beta) = 1 - (\abz' \cdot \abp)^2$, it follows
immediately from Eq. (\ref{app11}) that the $\hat{b}_p$ operators
satisfy the canonical commutation rules
\begin{equation}\label{app12}
\bigl[ \hat{b}_p,\hat{b}_p^\dagger \bigl] = 1.
\end{equation}

In order to clarify the relation between the quantum mechanical
relation Eq. (\ref{app11}) and the classical Jones matrix Eq.
(\ref{app7}), let us briefly summarize our procedure in a more
formal way.
 Let
$\abe^{(p)} \in \mathbb{R}^2$ be a given real unit vector
depending from a set of real variables shortly indicated with $p$.
As before, let $\aba^I$ and $\aba^T$ denote the operator vector
amplitudes of a plane-wave field before and after the polarizer,
respectively. Then we {\em define} the annihilation operator
$\hat{b}_p$ in the same way as in Eq. (\ref{1.4}), by imposing
\begin{equation}\label{app12_1}
\aba^I \cdot \abe^{(p)} = \aba^T \cdot \abe^{(p)} \equiv
\hat{b}_p.
\end{equation}
It is straightforward to show that:
\begin{equation}\label{app12_2}
\begin{array}{lcl}
  \bigl[ \hat{b}_p,\hat{b}_p^\dagger \bigl] & = & \hat{e}_\alpha^{(p)}\hat{e}_\beta^{(p)} \bigl[
  \hat{a}^T_\alpha, (\hat{a}^T_\beta)^\dagger \bigr]\\\\
   & = & \hat{e}_\alpha^{(p)}\hat{e}_\alpha^{(p)} = 1,\\
\end{array}
\end{equation}
where summation over repeated greek indices is understood.
Moreover, as in Eq. (\ref{linear})  we have defined
$\hat{a}^T_\alpha \equiv \aba_\bk^T \cdot \abe^{(\alpha)}_\bk$,
$(\alpha = 1,2)$, and Eq. (\ref{2.4}) has been used. Eq.
(\ref{app12_2}) suggests the introduction of a second rank
symmetric tensor $T^{p}$ defined as
\begin{equation}\label{app12_3}
 T^{p}_{\alpha \beta} \equiv  \hat{e}_\alpha^{(p)}\hat{e}_\beta^{(p)},
\end{equation}
such that $\mathrm{Tr}\, T^{p} = 1$. Now  let us define
$\abe^{(p)}$in terms of $\abp$ as in Eq. (\ref{app3})
\begin{equation}\label{app12_4}
D(p) \hat{e}^{(p)}_\alpha \equiv ( \delta_{\alpha \beta} -
\hat{k}_\alpha \hat{k}_\beta )\hat{p}_\beta,
\end{equation}
where $D^2(p) = ( \delta_{\alpha \beta} - \hat{k}_\alpha
\hat{k}_\beta ) \hat{p}_\alpha \hat{p}_\beta$. Then it is easy to
see that the tensor $T^{p}$ coincides with the Jones matrix $T$ in
Eq.(\ref{app7}).

From Eq. (\ref{app12}) follows that $\hat{b}_p$ is a genuine
bosonic operator, therefore we can use it, for different values of
$\beta$, to calculate the states of the field behind the
polarizer. In particular we define a one-photon state
$|\psi(\beta) \rangle$ as
\begin{equation}\label{app13}
\begin{array}{lcl}
|\psi(\beta) \rangle & = & \displaystyle{\hat{b}_p^\dagger
|0\rangle}
\\\\
& = & \displaystyle{\left( \frac{\abp \cdot \abx'
}{D(\beta)}\right) | x'(\abz') \rangle + \left( \frac{\abp \cdot
\aby' }{D(\beta)}\right) | y'(\abz') \rangle},
\end{array}
\end{equation}
where we have introduced the linear polarization basis states
defined as
\begin{equation}\label{app14}
\begin{array}{lcl}
| x'(\abz') \rangle & \equiv  & \displaystyle{\hat{a}^\dagger_{x'}
|0\rangle},
\\\\
| y'(\abz') \rangle & \equiv  & \displaystyle{\hat{a}^\dagger_{y'}
|0\rangle}.
\end{array}
\end{equation}
Moreover we define
\begin{equation}\label{app15}
\begin{array}{rclll}
|\psi(0) \rangle &\equiv & \hat{b}^\dagger_x | 0 \rangle & \equiv&
| x(\abz) \rangle, \\\\|\psi(\pi/2) \rangle &\equiv &
\hat{b}^\dagger_y | 0 \rangle &\equiv & | y(\abz) \rangle,
\end{array}
\end{equation}
where $\abz$ refers to the axis of the polarizer and {\em not} to
the photon momentum $\bk = \abz' |\bk|$. In fact, in a more
complete way, one should write, e.g., $|\bk, x(\abz)\rangle$ for
$| x(\abz)\rangle$. Clearly, Eq. (\ref{app15}) reduces to  Eq.
(\ref{app14}) when $\theta = 0 = \phi$. The two states Eq.
(\ref{app15}) have unit length but are not necessarily orthogonal;
in general we have
\begin{equation}\label{app16}
\langle \psi({\alpha}) | \psi(\beta) \rangle = \frac{\cos^2\theta
\cos \bar{\alpha} \cos \bar{\beta} + \sin \bar{\alpha} \sin
\bar{\beta} }{[\left(1 - \cos^2 \bar{\alpha} \sin^2 \theta \right)
\left(1 - \cos^2 \bar{\beta} \sin^2 \theta \right)]^{1/2}},
\end{equation}
where  $\bar{\alpha} = \alpha - \phi$, $\bar{\beta} = \beta -
\phi$. From Eq. (\ref{app16}) follows that $\langle \psi(\alpha) |
\psi(\beta) \rangle = 0$ when $\beta = \phi \pm \arctan(\cos
\theta)$ and $\alpha = 2 \phi + \pi - \beta$. In this case, when
assuming arbitrary $\beta$, the corresponding spatial orientations
$\abp(\alpha)$ and $\abp(\beta)$ are {\em not} orthogonal:
\begin{equation}\label{app17}
\abp(\alpha) \cdot \abp(\beta) = - \cos ( 2 \beta - 2 \phi),
\end{equation}
unless $\beta = \phi + \pi/4$. More generally, since $\phi$ is
arbitrary, we put $\phi = 0$ and we see that $\langle \psi(\beta +
\xi ) | \psi(\beta) \rangle = 0$ when
\begin{equation}\label{app16_1}
  \xi = \pi - \arccos \left[ \frac{ \sin \beta \cos \beta \sin^2 \theta}
  {\sqrt{1 - \cos^2 \beta \sin^2 \theta (1 + \cos^2 \theta)}}
  \right],
\end{equation}
which clearly reduces to $\pi/2$ when $\theta = 0$. Therefore we
conclude that is not possible to find a common orthogonal
polarization basis for all values of $\theta$.

Finally we can answer the question posed in the end of the
previous section. With the machinery we have built we can
calculate, for example, the probability amplitude that an
impinging photon in the state  $ | x'(\abz') \rangle$, is found
behind the polarizer in the state $ | x(\abz) \rangle$; the result
is:
\begin{equation}\label{app18}
 \langle  x(\abz)  | x'(\abz') \rangle  =  \frac{\abx \cdot
 \abx'}{D(0)}.
\end{equation}
More generally, we can calculate the {\em non-unitary}
transformation matrix $W(\bk)$ as
\begin{equation}\label{app19}
\begin{array}{lcl}
  W(\bk) & = &
\left(
\begin{array}{cc}
 \displaystyle{  \langle  x(\abz) | x'(\abz')\rangle} & \displaystyle{ \langle  x(\abz) | y'(\abz')
 \rangle}
  \\\\
   \displaystyle{\langle  y(\abz) | x'(\abz') \rangle } &  \displaystyle{ \langle  y(\abz) | y'(\abz')
   \rangle}
\end{array}
  \right)\\\\ & =&
 \left(
\begin{array}{cc}
 \frac{ \cos \phi \cos \theta}{\left(1 - \cos^2 \phi \sin^2 \theta\right)^{1/2}} &
   \frac{-\sin \phi}{\left(1 - \cos^2 \phi \sin^2
\theta\right)^{1/2}
}\\\\
\frac{ \sin \phi \cos \theta}{\left(1 - \sin^2 \phi \sin^2
\theta\right)^{1/2}} &
  \frac{\cos \phi}{\left(1 - \sin^2 \phi \sin^2 \theta\right)^{1/2}}
\end{array}
  \right).
  \end{array}
\end{equation}
The knowledge of $W(\bk)$ allows us to calculate all transition
amplitudes  between both linear and circular polarization states,
the latter being obtained by forming suitable linear combinations
of the elements $W_{ij}$. Note that $W$ is a real matrix because
we have chosen to represent a linear polarizer; it is possible to
show that in the case of an elliptic polarizer, complex phase
factors appear \cite{MandelBook}.
\subsection{Effective Reduced Density Matrix}
In Sec. II we have shown that it is not possible in general to
build a $2 \times 2$ reduced density matrix when many modes of the
field are involved. However, looking at Eq. (\ref{2.7}) one is
tempted to define an {\em average} $2 \times 2$ density matrix
$\bar{\rho}$ as
\begin{equation}\label{5.8}
\bar{\rho} \equiv \sum_{n=1}^N R(n).
\end{equation}
This seems a reasonable definition since in a real experiment, the
detectors automatically take averages over the photon momenta.
Unfortunately this choice works only if the measured quantities
{\em do not} depend on the momentum. In fact if, following the
same line of thinking as above, we define an average $2 \times 2$
matrix projector $\bar{P}$ as
\begin{equation}\label{5.9}
\bar{P} \equiv \frac{1}{N}\sum_{n=1}^N P(n),
\end{equation}
we can calculate its mean value as
\begin{equation}\label{4.7bis}
\bigl\langle {\bar{P}} \bigr\rangle = \mathrm{Tr}(\bar{\rho}
\bar{P}).
 \end{equation}
This definition coincide with the original one given by Eq.
(\ref{b70}) only in the special case  $P(n) = P(m) $ $\forall \,
n,m = 1, \dots, N$. This result  is hardly surprising, in fact by
comparing Eq. (\ref{b14}) with. Eq. (\ref{5.8}) one can easily
recognize that $\bar{\rho} = \rho^\mathrm{R}_1$.

Let us try now a different approach by exploiting the analogies
between classical and quantum optics. For a classical
quasi-monochromatic light wave propagating in the direction
$\abk$, a $2 \times 2$ polarization density matrix
$\rho_\mathrm{eff}$ can be defined in terms of the {\em measured}
Stokes parameters $s_i, (i = 0, \ldots,3)$ as
\cite{BornWolf,MandelBook}
\begin{equation}\label{A1}
\rho_\mathrm{eff} = \frac{1}{2} \left(
\begin{array}{cc}
  s_0 + s_3 & s_1 - i s_2 \\
  s_1+ i s_2 & s_0 - s_3
\end{array}
 \right).
\end{equation}
Clearly the procedure one adopts to measure the Stokes parameters
affects the actual value of $\rho_\mathrm{eff}$. Therefore, in
classical optics, the polarization density matrix
$\rho_\mathrm{eff}$ is understood as the {\em measured} density
matrix, and different measurement schemes will lead to different
matrices. Can we do the same in the quantum regime? As a matter of
fact, when we have a well defined experimental procedure to
measure the Stokes parameters of a beam of light, it is irrelevant
whether the beam contains $10^{20}$ or $1$ photon. Therefore we
regard the definition of $\rho_\mathrm{eff}$ as given by Eq.
(\ref{A1}), as a postulate valid both in the classical and in the
quantum regime.

The quantum theory of light gives us the rules to calculate the
Stokes parameters for both the one-photon \cite{JauchBook} and the
two-photon \cite{Abouraddy02} states. We consider here only the
one-photon case in some detail since the two-photon one is
completely analogous. For a given momentum $\bk$ and a polarizer
axis $\abz$, let $\mathcal{B}_\bk = \{ |\bk, x(\abz) \rangle,
|\bk, y (\abz)\rangle \}$ be the linear polarization basis defined
by two orthogonal polarizer orientation $\abp(0) \equiv \abx$,
$\abp(\pi/2) \equiv \aby$ introduced in the previous subsection.
When $\bk \parallel \abz$, in such a one-photon basis it is
possible to represent the ``Stokes operators'' $\hat{S}_i$
\cite{JauchBook} restricted to $\mathcal{H}_\bk$, by the
corresponding Pauli matrices: $\hat{S}_i | \mathcal{H}_\bk \doteq
\sigma_i$, $(i = 0, \ldots, 3)$, where, e.g.,
\begin{equation}\label{A3}
\begin{array}{lcl}
  \hat{S}_2| \mathcal{H}_\bk & = &  i \bigl( \hat{b}_y^\dagger \hat{b}_x - \hat{b}_x^\dagger \hat{b}_y
  \bigr)\\\\
   & \doteq & i \bigl(|y(\abz) \rangle \langle x(\abz)| - |x(\abz) \rangle \langle
   y(\abz)|)\\\\
   & \doteq & \left(
\begin{array}{cc}
  0 & -i \\
 i & 0
\end{array}
   \right)_\bk.
\end{array}
\end{equation}
When $\bk \nparallel \abz$, the Pauli matrices transform as
$\sigma_i \rightarrow W^T \sigma_i W$. The mean values $s_i =
\langle \hat{S}_i \rangle$ can be calculated by using the {\em
total} scattering density matrix $\hat{\rho}_f \equiv | \psi_f
\rangle \langle \psi_f |$ as
\begin{equation}\label{A4}
s_i = \mathrm{Tr} (\hat{\rho}_f \hat{S}_i), \qquad (i = 0, \ldots,
3 ),
\end{equation}
where the state $| \psi_f \rangle$ is given in Eq. (\ref{5.4}).
Looking at Eq. (\ref{A3}) it is clear that all we have to
calculate are the mean values of the four operators
$\hat{P}_{\sigma \tau}$ $(\sigma, \tau = x,y)$ defined as
\begin{equation}\label{A5}
\hat{P}_{\sigma \tau} = \sum_{\bk \in K_D} | \bk, \sigma(\abz)
\rangle \langle \bk, \tau(\abz)|.
\end{equation}
Note that the off-diagonal operators $\hat{P}_{\sigma  \tau}$
($\sigma \neq \tau $) do not correspond to physical observables,
therefore  are not Hermitian. Finally, after comparing Eqs.
(\ref{A1}-\ref{A4}) and Eq. (\ref{A5}), we can write
\begin{equation}\label{A7}
\bigl( \rho_\mathrm{eff} \bigr)_{\sigma \tau} =Z_{\mathrm{eff}}
\bigl\langle \hat{P}_{\tau \sigma} \bigr\rangle,
\end{equation}
where $Z_\mathrm{eff} = 1/(\langle \hat{P}_{xx} \rangle + \langle
\hat{P}_{yy} \rangle)$ is a normalization constant which ensures
$\mathrm{Tr} \rho_\mathrm{eff} = 1$.  Explicitly we have
\begin{equation}\label{A8}
\frac{\rho_\mathrm{eff}}{Z_\mathrm{eff}} = \left(
\begin{array}{cc}
  \langle \hat{P}_{xx} \rangle & \langle \hat{P}_{yx} \rangle \\\\
  \langle \hat{P}_{xy} \rangle & \langle \hat{P}_{yy} \rangle
\end{array}
\right).
\end{equation}
This step completes our calculation.
 The presence of the normalization constant
$Z_{\mathrm{eff}}$ should not be surprising, it simply amounts to
a renormalization of the Stokes parameters $s_i$ with respect to
$s_0$: $s_i \rightarrow s_i/s_0$.

Without repeating all the calculations, we shall give in the next
Section directly the formula corresponding to Eq. (\ref{A7}) for
the two-photon case.

\section{Two-Photon Scattering and the Bell-CHSH Inequalities}
\subsection{Two-Photon Scattering Matrix}
We consider now the following experimental configuration. A
two-photon source emits a pair of polarization-entangled photons
\cite{Kwiat95} and sends them through two scattering systems $S_A$
and $S_B$ located along the photon paths. Two linear polarizers
$P_A$ and $P_B$ are put in front of two multi-mode detectors $D_A$
and $D_B$ which can record both the two singles count rates and
the coincidence count rate.

The two-photon initial state emitted by the source is the Bell
state
\begin{equation}\label{Last1}
| \psi_i \rangle =
C_+|\psi_A^x\rangle|\psi_B^y\rangle+C_-|\psi_A^y\rangle|\psi_B^x\rangle,
\end{equation}
where $C_\pm$ are complex coefficients such that $|C_+|^2 +
|C_-|^2 =1 $, the subscripts $A$ and $B$ identify the two photons,
and the superscripts $x$ and $y$ denote the {\em linear}
polarization state. Moreover
\begin{equation}\label{Last2}
|\psi_F^\alpha\rangle \equiv |\bk_F,
\alpha(\bk_F)\rangle_{\mathrm{in}}, \qquad (\alpha=x,y; \; F =
A,B).
\end{equation}
The state $| \psi_i\rangle$ is an eigenstate of the total linear
momentum operator $\hat{\mathbf{K}}_{AB} = \hat{\mathbf{K}}_A
\otimes \hat{\mathbf{1}}_B + \hat{\mathbf{1}}_A  \otimes
\hat{\mathbf{K}}_B$:
\begin{equation}\label{Last2bis}
\hat{\mathbf{K}}_{AB}|\psi_i \rangle = (\bk_A + \bk_B)|\psi_i
\rangle.
\end{equation}
 So, at this stage, it is still
possible to describe the state $| \psi_i \rangle$ in terms of a $4
\times 4$ ``polarization part'' density matrix.

The state $|\psi_f\rangle$ of the pair after the scattering has
occurred, can be written as
\begin{equation}\label{Last3}
|\psi_f \rangle = \sum_{ \bk \in K \atop \bq \in Q }\sum_{ \alpha,
\beta} \Psi_{\alpha \beta}(\bk, \bq)
 | \bk, \alpha(\bk) \rangle_A | \bq, \beta(\bq) \rangle_B,
\end{equation}
where
\begin{equation}\label{Last4}
\Psi_{\alpha \beta}(\bk, \bq) \equiv C_+ S_{\alpha x}^A S_{\beta
y}^B + C_- S_{\alpha y}^A S_{\beta x}^B,
\end{equation}
and
\begin{equation}\label{Last5}
\begin{array}{ccc}
S_{\alpha \xi}^A & \equiv &\langle \bk, \alpha(\bk)| \bk_A,
\xi(\bk_A)\rangle_{\mathrm{in}},\quad (\xi = x,y),\\\\
S_{\beta \eta}^B & \equiv &\langle \bq, \beta(\bq)| \bk_B,
\eta(\bk_B)\rangle_{\mathrm{in}}, \quad (\eta = x,y),
\end{array}
\end{equation}
are the scattering matrix elements \cite{KakuBook}. Moreover, $K$
and $Q$ denote the sets of all scattered modes for photon $A$ and
$B$ respectively, and  $\alpha = \alpha(\bk)$,  $\beta =
\beta(\bq)$. By inspecting Eq. (\ref{Last3}) it is easy to see
that the state $| \psi_f \rangle$ is not longer an eigenstate of
the linear momentum. Now, by repeating the same procedure we have
executed for the one-photon scattering case, we introduce the
two-photon generalized reduced density matrix as
\begin{equation}\label{Last6}
\hat{\rho}^\mathrm{R}_2  = \sum_{ \bk\in K_{D_A}\atop \bq
 \in Q_{D_B} } w(\bk, \bq)
| \phi (\bk, \bq) \rangle \langle \phi (\bk, \bq) | ,
\end{equation}
where we have defined
\begin{equation}\label{Last6.1}
| \phi (\bk, \bq) \rangle  = \frac{1}{\sqrt{\zeta(\bk,\bq)}}
\sum_{\alpha, \beta } \Psi_{\alpha \beta}(\bk, \bq)| \bk,
\alpha(\bk) \rangle_A | \bq, \beta(\bq) \rangle_B,
\end{equation}
where $K_{D_A}$, $Q_{D_B}$ represent the sets of the scattered
modes detected by detectors $D_A$ and $D_B$ respectively.
Moreover, we have defined
\begin{equation}\label{Last6.2}
\begin{array}{rcl}
  \displaystyle{\zeta(\bk, \bq)}  & = & \displaystyle{\sum_{\alpha, \beta}|\Psi_{\alpha\beta}(\bk,\bq)|^2} ,\\\\
  \displaystyle{Z_2}  & = & \displaystyle{1/\displaystyle{\sum_{ \bk\in K_{D_A}\atop \bq
 \in Q_{D_B} }} \zeta(\bk, \bq)}  ,\\\\
  \displaystyle{w(\bk, \bq)}  & = & \displaystyle{Z_2 \zeta(\bk, \bq)}.
\end{array}
\end{equation}
As in the one-photon case, the operation of tracing with respect
to the detected modes has reduced the {\em pure} state $| \psi_f
\rangle$ to the statistical {\em mixture}
$\hat{\rho}^\mathrm{R}_2$.
Now it is clear that we can introduce a set of $N_A N_B$'s  pure
state density matrices $\tilde{{\rho}}(\bk,\bq)$ ($4 \times 4$)
whose elements are
\begin{equation}\label{Last7}
\tilde{\rho}_{\alpha \beta, \alpha' \beta'}(\bk,\bq) =
\frac{\Psi_{\alpha \beta}(\bk, \bq) \Psi_{\alpha' \beta'}^*(\bk,
\bq)}{\zeta(\bk,\bq)},
\end{equation}
where $N_A = \dim K_{D_A}$, $N_B = \dim Q_{D_B}$. Each of these
matrices ``lives'' in a 4-dimensional  Hilbert space
$\mathcal{H}_{\bk \bq}$:
\begin{equation}\label{Last8}
\mathcal{H}_{\bk \bq} = \mathrm{span} \{ |\bk,
\alpha(\bk)\rangle_A \otimes  |\bq, \beta(\bq)\rangle_B \},
\end{equation}
where $(\bk \in K_{D_A}, \bq \in Q_{D_B})$ and $(\alpha,\beta =
x,y)$.

Not surprisingly, we have found a result analogous the one-photon
scattering case, that is that a {\em unique} reduced $4 \times 4$
density matrix does not exist. However, by using the methods
developed in Sec. III we can introduce an effective $4 \times 4$
reduced density matrix $\rho_{2 \mathrm{eff}}$.
In analogy with the one-photon case, $\rho_{2 \mathrm{eff}}$ can
be written in terms of the measured two-photon Stokes parameters
$s_{ij}$ ($i,j=0,\ldots,3$) \cite{Abouraddy02} as:
\begin{equation}\label{A8bis}
\rho_{2 \mathrm{eff}} = \frac{1}{4} \sum_{i,j=0}^3  s_{ij}
(\sigma_i \otimes \sigma_j).
\end{equation}
Then, by following the same line of reasoning of the one-photon
case, one can realize that it is possible to write
\begin{equation}\label{A9}
\bigl( \rho_{2 \mathrm{eff}} \bigr)_{\alpha \beta, \alpha' \beta'}
= Z_{2 \mathrm{eff}} \bigl\langle \hat{P}_{\alpha' \alpha} \otimes
\hat{P}_{\beta' \beta} \bigr\rangle,
\end{equation}
where
\begin{equation}\label{A10}
\begin{array}{lcl}
\hat{P}_{\alpha' \alpha} & = & \displaystyle{\sum_{\bk \in K_D}
|\bk, \alpha'(\abz_A) \rangle \langle \bk,
\alpha(\abz_A)|, \qquad (\alpha, \alpha' = x,y) },\\\\
 \hat{P}_{\beta' \beta} & = &\displaystyle{\sum_{\bq \in Q_D} |\bq, \beta'(\abz_B) \rangle \langle \bq,
\beta(\abz_B)|,\qquad (\beta, \beta' = x,y) },
\end{array}
\end{equation}
and $Z_{2 \mathrm{eff}}$ such that $\mathrm{Tr} \rho_{2
\mathrm{eff}}=1$. With $\abz_A$ and $\abz_B$ we have denoted the
axes of the two polarizers located in the paths of the photons $A$
and $B$ respectively.

The calculation of $\rho_{2 \mathrm{eff}}$ we have sketched above,
closely resembles the previous one for the one-photon case. There
is, however, an important conceptual difference between the two
cases, as emphasized in Ref. \cite{Abouraddy02}. In fact, in the
two-photon case  $\rho_{2 \mathrm{eff}}$  cannot be determined by
local measurement only (each beam separately), but it is necessary
to make {\em coincidence} measurement in order
 to account for the (possible) entanglement between the two photons.
However, it is well known that entanglement properties of a
bipartite system depend on the dimensionality of the underlying
Hilbert space \cite{Collins02,Law00}, therefore the
``measurement-induced''
 reduction from $4 \times N_A \times N_B$ to $4$
dimensions, may change the observed properties of the system. The
problem of the determination of the effective dimensionality  of
the scattered pair state, is at present under investigation in our
group \cite{Aiello04_2}.
\subsection{The Bell-CHSH Inequality}
We have just shown that, when in a two-photon scattering process
we have a multi-mode detection scheme, the  polarization state of
the photon pair is reduced to a statistical mixture.
We want to study the violation of the Bell inequality in the
Clauser, Horne, Shimony and Holt (CHSH) form \cite{Clauser69}, for
that mixture. As usual the Bell operator
$\hat{\mathcal{B}}_{\mathrm{CHSH}}$ is defined as
\cite{Braunstein92}
\begin{equation}\label{Last10}
\hat{\mathcal{B}}_{\mathrm{CHSH}} = \aba \cdot {\bm {\sigma }}
\otimes (\abb + \abb')\cdot \bm{\sigma} + \aba' \cdot {\bm {\sigma
}} \otimes (\abb - \abb')\cdot \bm{\sigma},
\end{equation}
where  $\aba, \aba', \abb$ and $\abb'$ are unit vectors in
$\mathbb{R}^3$. Moreover, $\bm{\sigma}$ is a vector built with the
three standard Pauli matrices $\sigma_1, \sigma_2, \sigma_3$, and
the scalar product $\aba \cdot {\bm {\sigma }}$ stand for the $2
\times 2$ matrix $\sum_{i=1}^3 a_i \sigma_i $.
 Then the CHSH inequality is
\begin{equation}\label{Last20}
  \bigl| \mathrm{Tr} ( \hat{\rho}\hat{\mathcal{B}}_{\mathrm{CHSH}}) \bigr| \leq 2.
\end{equation}
In order to calculate explicitly Eq. (\ref{Last20}) it is
necessary to know $\hat{\rho}$ which, in turn, depends on the
specific scattering process considered. However, in our case, we
want to show that the polarization-entanglement of a photon pair
is  degraded just because of the multi-mode detection, {\em
independently} from the details of the process; therefore we shall
consider a very general shape for $\hat{\rho}$.

 Let
$w(\bk,\bq) \geq 0$ denote the probability  of a given physical
realization of the process. Then we can write
\begin{equation}\label{Last21}
\hat{\rho} = \sum_{\bk,\bq}w(\bk,\bq)  |\Psi_{\bk \bq} \rangle
\langle \Psi_{\bk \bq}|,
\end{equation}
where $ |\Psi_{\bk \bq} \rangle$ represent an arbitrary
polarization entangled state for which the photons $A$ and $B$
have momenta $\bk$ and $\bq$ respectively.
This means that each time a pair is scattered, both photons will
impinge on the corresponding polarizers with different angles
determined by their momenta. So, for our purposes is enough to
investigate the angular dependence of the entanglement of a {\em
single} emitted photon pair, when  at least one of the two photons
impinges with an arbitrary angle on the corresponding polarizer.
\begin{figure}[!ht]
\includegraphics[angle=0,width=8truecm]{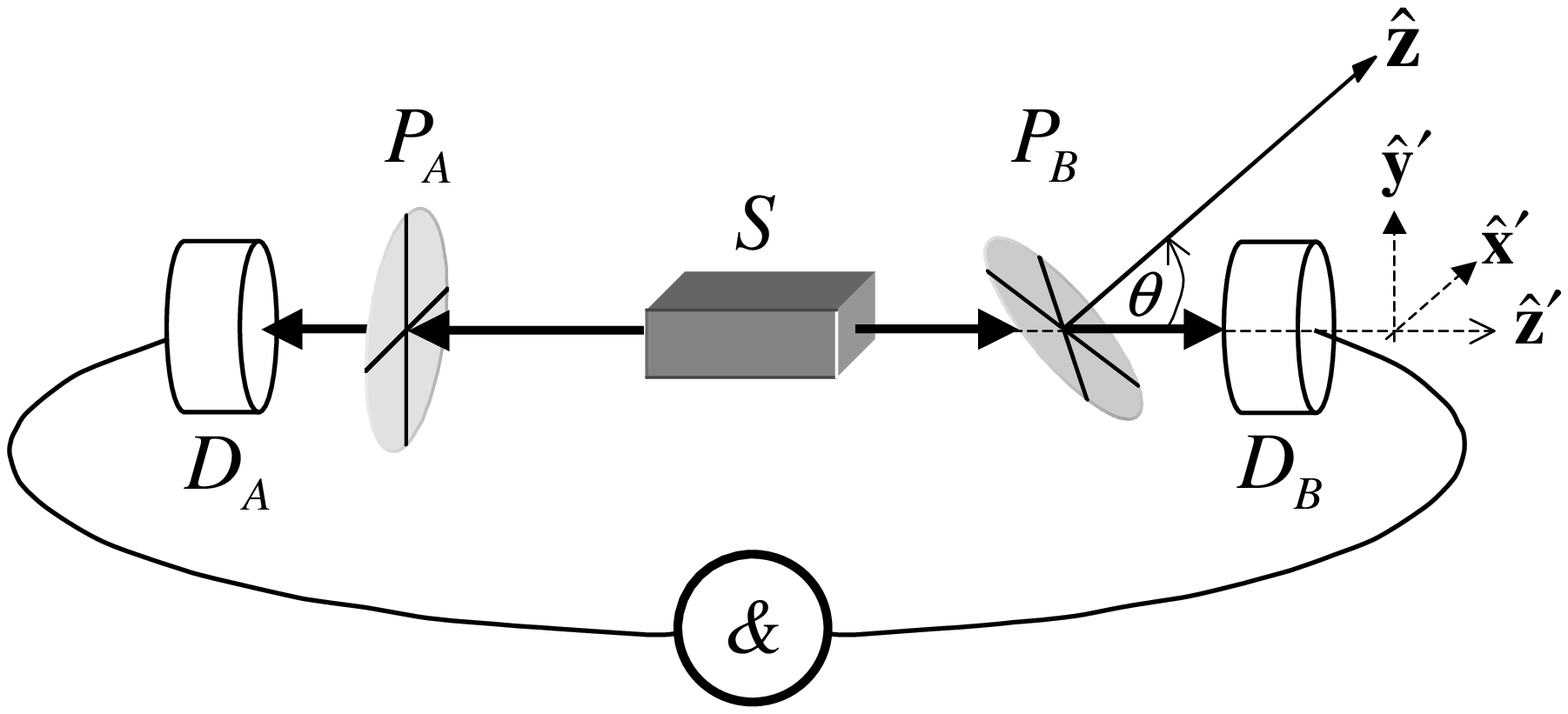}
\caption{\label{fig:1}  Scheme of the two-photon analyzer
described in the text. $S$ is a source of Bell-Schmidt states,
$P_A$ and $P_B$ are linear polarizers, and $D_A$ and $D_B$ are
photo-detectors. The symbol \& denotes the coincidence recorder.}
\end{figure}
In order to keep our treatment as general as possible, instead of
considering a particular process-dependent scattered state, we
focus our attention on the entanglement properties of the complete
set provided by the Bell-Schmidt states \cite{Aravind95}
\begin{equation}\label{Last30}
\begin{array}{ccc}
  |\Phi_1 \rangle & = & \frac{1}{\sqrt{2}} (| x_A, x_B \rangle + | y_A, y_B \rangle)
  ,\\\\
  |\Phi_2 \rangle & = & \frac{1}{\sqrt{2}} (| x_A, x_B  \rangle - | y_A, y_B \rangle),
  \\\\
  |\Phi_3 \rangle & = & \frac{1}{\sqrt{2}} (| x_A, y_B \rangle + | y_A, x_B \rangle),
  \\\\
  |\Phi_4 \rangle & = & \frac{1}{\sqrt{2}} (| x_A, y_B \rangle - | y_A, x_B
  \rangle).
  \\
\end{array}
\end{equation}
Then we associate to each Bell-Schmidt state a well defined photon
momentum pair $(\bk_A, \bk_B)$ and we show that for each pure
state $| \Phi_i\rangle$, the optimal choices of $\aba, \aba',
\abb$ and $\abb'$ depend on $(\bk_A, \bk_B)$ and therefore it is
impossible to find a choice which is simultaneously optimal for
all the states in the ensemble given in Eq. (\ref{Last21}).

In order to demonstrate this, let us consider the detection
coincidence scheme shown in Fig. 1. An idealized source $S$ emits
photon pairs  in  the Bell-Schmidt states $| \Phi_i\rangle$. Two
linear polarizers $P_A$ and $P_B$ are inserted in the paths of the
two photons and two detectors $D_A$ and $D_B$ are put behind them.
While $P_A$ is put perpendicular to the momentum $\bk_A$ of the
photon $A$, the axis $\abz$ of $P_B$ is such that $\abz \cdot
\bk_B /|\bk_B|=\cos \theta$.

Aravind \cite{Aravind95} has shown that the choices $a_z = 1$,
$a'_x=1$, $b_y = 0 =b_y'$ are optimal for all the $|
\Phi_i\rangle$, therefore we make the same assumptions. The
remaining components of the two vectors $\abb$ and $\abb'$ can be
related to the {\em physical} orientations of the polarizer by
writing,
\begin{equation}\label{Last40}
\begin{array}{lcl}
\abr & = & \mathrm{Tr}(T \bm{\sigma})\\\\
   & = & \bigl\{ \mathrm{Tr}(T \sigma_1), \; \;  0, \;  \; \mathrm{Tr}(T \sigma_3)
   \bigr\},
\end{array}
\end{equation}
where $\abr = \abb, \abb'$.  $T$ is the polarizer Jones matrix as
given in Eq. (\ref{app7}) and $ \mathrm{Tr}(T \sigma_2)=0$ because
of the symmetry of $T$. Then we parameterize $\abb$ and $\abb'$ by
introducing the two angles $\beta$ and $\delta$ respectively, in
the Eq. (\ref{Last40}) obtaining
\begin{equation}\label{Last50}
\begin{array}{lcl}
\abb & = & \displaystyle{\left\{
  \frac{2 \cos \theta \sin  {\beta}  \cos  {\beta}  }{1 - \cos^2  {\beta}  \sin^2\theta},
   0,
  \frac{ \cos^2 \theta \cos^2  {\beta}  -\sin^2  {\beta}  }{1 - \cos^2  {\beta}  \sin^2\theta}  \right\}},\\\\
\abb' & = & \displaystyle{\left\{
  \frac{2 \cos \theta \sin  {\delta}  \cos  {\delta}  }{1 - \cos^2  {\delta}  \sin^2\theta},
   0,
  \frac{ \cos^2 \theta \cos^2  {\delta}  -\sin^2  {\delta}  }{1 - \cos^2  {\delta}  \sin^2\theta}
  \right\}},\\
\end{array}
\end{equation}
where ${\beta}$ stands for  $\beta - \phi$ and ${\delta}$ for
$\delta - \phi$. Now for each of the Bell-Schmidt states Eq.
(\ref{Last30}) we choose the values for ${\beta}$ and ${\delta}$
in order to maximize the violation of the Bell-CHSH inequality for
$\theta = 0$, and calculate
\begin{equation}\label{Last60}
B_i(\theta) \equiv \mathrm{Tr} \bigl( \hat{\rho}_i
\hat{\mathcal{B}}_{\mathrm{CHSH}} \bigr), \qquad (i = 1,\ldots,4),
\end{equation}
where we have defined $\hat{\rho}_i = | \Phi_i \rangle \langle
\Phi_i |$. After a straightforward calculations one finds that
$B_1(\theta) = B_2(\theta)$ and $B_3(\theta) = B_4(\theta)$, where
\begin{equation}\label{Last70}
\begin{array}{lcl}
  B_2(\theta) & = & \displaystyle{-4\cos^2\left( \frac{\theta}{2} \right) \frac{1 - (3 + 2 \sqrt{2})\cos \theta}
  {1 + (3 + 2 \sqrt{2})\cos^2 \theta}}, \\\\
  B_4(\theta) & = &  \displaystyle{4\cos^2\left( \frac{\theta}{2} \right) \frac{1 - (3 - 2 \sqrt{2})\cos \theta}
  {1 + (3 - 2 \sqrt{2})\cos^2 \theta}}.
\end{array}
\end{equation}
These functions are plotted in Fig. 2.
\begin{figure}[!ht]
\includegraphics[angle=0,width=9truecm]{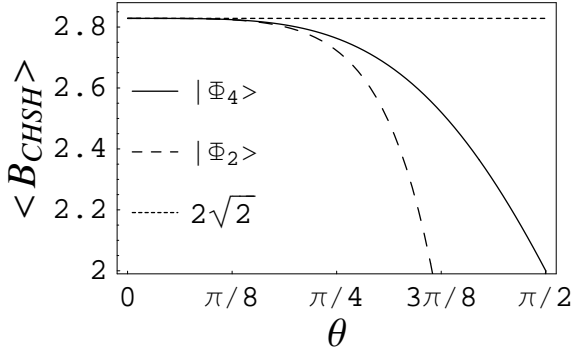}
\caption{\label{fig:2}  Average Bell-CHSH operator $\langle
\hat{\mathcal{B}}_{\mathrm{CHSH}}\rangle$ as a function of the
polarizer ``tilting'' angle $\theta = \arccos (\abz \cdot
\abk_B)$,
 for the Bell-Schmidt states $|\Phi_4\rangle$ and
$|\Phi_2\rangle$. The unit vectors $\abb$ and $\abb'$ have been
chosen such that for both states $\langle
\hat{\mathcal{B}}_{\mathrm{CHSH}}\rangle = 2 \sqrt{2}$ for $\theta
= 0$. When $\theta$ increases, the violation of the Bell-CHSH
inequality decreases. The same curves have been obtained for
$|\Phi_1\rangle$ (alike $|\Phi_2\rangle$) and $|\Phi_3\rangle$,
(alike $|\Phi_4\rangle$).}
\end{figure}
It is clear that the optimal choice $(\beta^\mathrm{opt}_0,
\delta^\mathrm{opt}_0)$ at $\theta = 0$, is no longer valid when
$\theta$ increases and the degree of entanglement of skew photons
appears to be reduced. However, one must realize that this loss of
entanglement is an artefact due to our mismatched polarization
detector. This means that it is still possible to find optimal
values for $(\beta, \delta)$, but they will differ from the
initial ones ($\theta = 0$ case). In order to show this
explicitly, we have investigated the dynamics of points
$(\beta^\mathrm{opt}, \delta^\mathrm{opt})$ in the plane $(\beta,
\delta)$, for varying $\theta$. The results are shown in Fig. 3 in
the case of $| \Phi_4 \rangle$: for the other states the results
are qualitatively similar.
\begin{figure}[!ht]
\includegraphics[angle=0,width=9truecm]{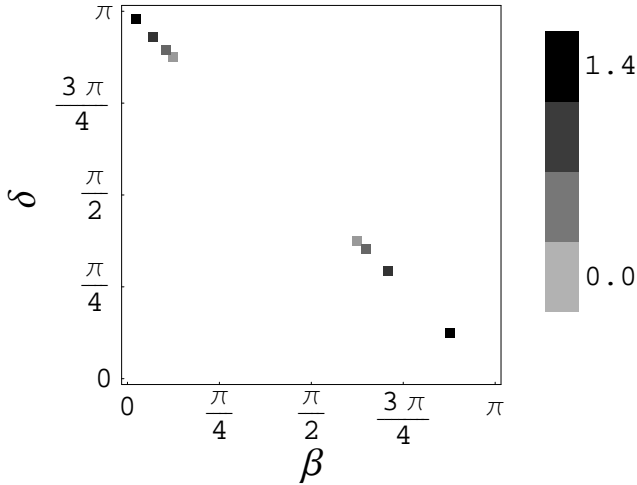}
\caption{\label{fig:3} The square boxes represent points in the
plane $(\beta, \delta)$ such that $\langle
\hat{\mathcal{B}}_{\mathrm{CHSH}}\rangle = \pm 2 \sqrt{2}$ ($+
\Leftrightarrow $ lower series, $- \Leftrightarrow$ upper series)
when the average is calculated with respect to $| \Phi_4 \rangle$
for several values of $\theta$. When $\theta$ increases passing
from zero (light grey) to about $\pi/2$ (dark grey), the ``maximal
violation points'' move monotonically from light to dark along the
line $\delta = \pi - \beta$.}
\end{figure}
When $\theta$ increases passing from zero  to $\pi/2$ , the points
$(\beta^\mathrm{opt}, \delta^\mathrm{opt})$ move monotonically
away from the central point $(\pi/2,\pi/2)$ along the line $\delta
= \pi - \beta$ with different rates.
\begin{figure}[!hb]
\includegraphics[angle=0,width=9truecm]{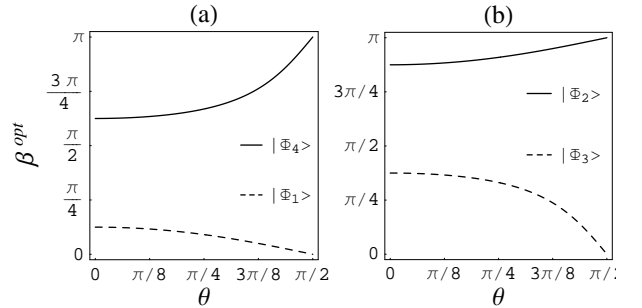}
\caption{\label{fig:4} (a-b) Dependence of the optimal value of
$\beta$ (such that $\langle
\hat{\mathcal{B}}_{\mathrm{CHSH}}\rangle =  2 \sqrt{2}$), by the
tilting angle $\theta$. For $\theta \gtrsim \pi/4$ different
values of $\theta$ may require quite different values of $\beta$.
}
\end{figure}
Once $\delta$ is fixed to the optimal value $\delta^\mathrm{opt} =
\pi - \beta$, one can follow the motion of $\beta^\mathrm{opt}$ as
a function of $\theta$. The dynamics is very simple and it is
shown in Fig. 4. We list below the four functions
$\beta^\mathrm{opt}_i(\theta)$, $(i = 1,\dots,4)$ for reference:
\begin{equation}\label{Last80}
\begin{array}{lcl}
  \beta^\mathrm{opt}_1(\theta) & = & - \arctan\left[(1 - \sqrt{2} \cos \theta)
  \right],
  \\\\
  \beta^\mathrm{opt}_2(\theta) & = & \pi + \arctan\left[(1 - \sqrt{2} \cos \theta)
  \right],
  \\\\
  \beta^\mathrm{opt}_3(\theta) & = &  \arctan\left[(1 + \sqrt{2} \cos \theta)
  \right],
  \\\\
  \beta^\mathrm{opt}_4(\theta) & = & \pi - \arctan\left[(1 + \sqrt{2} \cos \theta)
  \right].\\
\end{array}
\end{equation}
Despite  their simplicity, Fig. 4 and Eqs. (\ref{Last80}) tell us
something important. We remind that the idealized experimental
scheme we have considered in Fig. 1 was introduced to study the
behavior of an entangled photon-pair in a statistical mixture in
which each photon-pair has a well defined momentum. While in the
above analysis we have considered $\theta$ as a {\em free}
parameter representing the polarizer axis, in a real scattering
experiment $\theta$ is the angle at which one of the photons,
belonging to the entangled pair, impinges on the polarization
detector. Then, each time a pair is scattered the two photons $A$
and $B$ will hit the detectors with  arbitrary angles $\theta_A$
and $\theta_B$ respectively and the optimal polarizer orientation
$\beta^\mathrm{opt}$ will be different for each couple of angles.
Therefore it is clear that we cannot simultaneously optimize
$\beta$ for all angles and the measured {\em average} degree of
entanglement will be reduced independently from the scattering
process considered. This completes our proof.
Then we conclude that a conventional experimental setup for the
measurement of the Bell-CHSH inequality, may fail to give the
correct value for $\langle
\hat{\mathcal{B}}_{\mathrm{CHSH}}\rangle$ when the measured state
is a multi-mode scattered state.
\section{Summary and Outlook}
The present paper aims to establish a theoretical background  for
a future study of scattering processes by both chaotic optical
devices \cite{Aiello03} and random media. The main concern of this
paper has been to demonstrate that in a scattering process the
measured degree of polarization-entanglement of a photon pair is
unavoidably decreased because of the multi-mode detection. At the
core of this loss of entanglement process resides the correlation
between the momentum degrees of freedom and the polarization ones
for the states of the electromagnetic field. In order to clarify
the meaning of this correlation (or entanglement
\cite{PeresBook}), we have developed early on the paper a proper
notation for the representation of the one-photon states of the
electromagnetic field; this notation plays an important role
throughout the paper. This introductory part also serves as a
basis to show which dangers may be hidden behind the use of a
misleading notation. In particular we show that the use of a
reduced density matrix obtained by blindly tracing out the
momentum degrees of freedom, can lead to a wrong result when
applied to the calculation  of a polarization-dependent
observable.

The central part of this paper  comprises two  separate topics.
The first topic  consists in a careful analysis of the one-photon
scattering processes. It is shown that a unique $2 \times 2$
reduced density matrix is an useless concept for the analysis of a
multi-mode scattering process, and that more information than this
is required. The second topic is  how to build, within the QED
context, the one-photon states selected by an arbitrarily oriented
linear polarizer. The knowledge of these states allows us to
introduce the concept of the {\em effective} reduced density
matrix which must be understood as the {\em measured} density
matrix.

The last part of this paper is devoted to a brief introduction to
the subject of the two-photon scattering processes and to the
investigation of the Bell-CHSH inequality when, in a standard
measurement setup, a polarization analyzer is arbitrarily tilted.
The violation of the Bell-CHSH inequality is explicitly calculated
for the complete Bell-Schmidt set of polarization-entangled
states. We show that, when in a two-photon scattering experiment,
the observer is ignorant about the momentum distribution of the
scattered photons,  he cannot find an optimal orientation for the
polarizers  in order to maximize the {\em measured} violation of
the Bell-CHSH inequality. However this does not mean that a
scattering process necessarily spoils the degree of entanglement
of a given state, but instead just makes it not measurable with a
standard measurement setup. This naturally raises a question about
the physical meaning of a {\em computable} degree of entanglement
which does not coincide with the {\em measurable} one. This topic
is currently under investigation in our group \cite{Aiello04_2}.

\begin{acknowledgments}
The authors have greatly benefitted from many discussions with
Cyriaque Genet who is warmly acknowledged. We also have had
insightful discussions with Graciana Puentes who is acknowledged.
This work is supported by the EU under the IST-ATESIT contract and
also  by FOM.
\end{acknowledgments}


\begin{thebibliography}{29}
\expandafter\ifx\csname
natexlab\endcsname\relax\def\natexlab#1{#1}\fi
\expandafter\ifx\csname bibnamefont\endcsname\relax
  \def\bibnamefont#1{#1}\fi
\expandafter\ifx\csname bibfnamefont\endcsname\relax
  \def\bibfnamefont#1{#1}\fi
\expandafter\ifx\csname citenamefont\endcsname\relax
  \def\citenamefont#1{#1}\fi
\expandafter\ifx\csname url\endcsname\relax
  \def\url#1{\texttt{#1}}\fi
\expandafter\ifx\csname
urlprefix\endcsname\relax\def\urlprefix{URL }\fi
\providecommand{\bibinfo}[2]{#2}
\providecommand{\eprint}[2][]{\url{#2}}

\bibitem[{\citenamefont{Ballentine}(1986)}]{Ballentine}
\bibinfo{author}{\bibfnamefont{L.~E.} \bibnamefont{Ballentine}},
  \bibinfo{journal}{Am. J. Phys.} \textbf{\bibinfo{volume}{55}},
  \bibinfo{pages}{785} (\bibinfo{year}{1986}), \bibinfo{note}{and references
  therein}.

\bibitem[{\citenamefont{Nielsen and Chuang}(2002)}]{NielsenBook}
\bibinfo{author}{\bibfnamefont{M.~A.} \bibnamefont{Nielsen}} \bibnamefont{and}
  \bibinfo{author}{\bibfnamefont{I.~L.} \bibnamefont{Chuang}},
  \emph{\bibinfo{title}{Quantum Computation and Quantum Information}}
  (\bibinfo{publisher}{Cambridge University Press},
  \bibinfo{address}{Cambridge, UK}, \bibinfo{year}{2002}),
  \bibinfo{edition}{reprinted first} ed.

\bibitem[{\citenamefont{Zeilinger}(1999)}]{Zeilinger99}
\bibinfo{author}{\bibfnamefont{A.}~\bibnamefont{Zeilinger}},
  \bibinfo{journal}{Rev. Mod. Phys.} \textbf{\bibinfo{volume}{71}},
  \bibinfo{pages}{S288} (\bibinfo{year}{1999}).

\bibitem[{\citenamefont{Yariv}(1989)}]{YarivBook}
\bibinfo{author}{\bibfnamefont{A.}~\bibnamefont{Yariv}},
  \emph{\bibinfo{title}{Quantum Electronics}} (\bibinfo{publisher}{Johon Wiley
  \& Son, New York}, \bibinfo{year}{1989}), \bibinfo{edition}{3rd} ed.

\bibitem[{\citenamefont{Gisin et~al.}(2002)\citenamefont{Gisin, Ribody, Tittel,
  and Zbinden}}]{Gisin02}
\bibinfo{author}{\bibfnamefont{N.}~\bibnamefont{Gisin}},
  \bibinfo{author}{\bibfnamefont{G.}~\bibnamefont{Ribody}},
  \bibinfo{author}{\bibfnamefont{W.}~\bibnamefont{Tittel}}, \bibnamefont{and}
  \bibinfo{author}{\bibfnamefont{H.}~\bibnamefont{Zbinden}},
  \bibinfo{journal}{Rev. Mod. Phys.} \textbf{\bibinfo{volume}{74}},
  \bibinfo{pages}{145} (\bibinfo{year}{2002}).

\bibitem[{\citenamefont{Altewischer et~al.}(2002)\citenamefont{Altewischer, van
  Exter, and Woerdman}}]{Altewischer02}
\bibinfo{author}{\bibfnamefont{E.}~\bibnamefont{Altewischer}},
  \bibinfo{author}{\bibfnamefont{M.~P.} \bibnamefont{van Exter}},
  \bibnamefont{and} \bibinfo{author}{\bibfnamefont{J.~P.}
  \bibnamefont{Woerdman}}, \bibinfo{journal}{Nature}
  \textbf{\bibinfo{volume}{418}}, \bibinfo{pages}{304} (\bibinfo{year}{2002}).

\bibitem[{Sca()}]{Scatheory}
\bibinfo{note}{C. Genet, M. P. van Exter and J. P. Woerdman, Opt. Commun. {\bf
  225}, 331 (2003); J. L. van Velsen, J. Twarzydlo, and C. W. J. Beenakker,
  Phys. Rev. A {\bf 68}, 043807 (2003); E. Moreno, F. J. Garc\'ia-Vidal, D.
  Erni, J. I. Cirac, and L. Mart\'in-Moreno, quant-ph/0308075; C. Genet, E.
  Altewischer, M. P. van Exter and J. P. Woerdman, quant-ph/0311137}.

\bibitem[{Vel()}]{Velsen04}
\bibinfo{note}{J. L. van Velsen and C. W. Beenakker, quant-ph/0403093}.

\bibitem[{Woe()}]{Woerdman03}
\bibinfo{note}{J. P. Woerdman, talk presented at the Workshop {\em Fundamentals
  of Solid State Quantum Information Processing}, Lorentz Center, Leiden
  (2003)}.

\bibitem[{Per()}]{Peres_et_al}
\bibinfo{note}{A. Peres, and D. R. Terno, J. Mod. Opt. {\bf 50}, 1165 (2003);
  N. H. Lindner, A. Peres, and D. R. Terno, J. Phys. A {\bf 36}, L449 (2003);
  A. Peres, and D. R. Terno, Rev. Mod. Phys. {\bf 76}, 93 (2004)}.

\bibitem[{\citenamefont{Peres}(1998)}]{PeresBook}
\bibinfo{author}{\bibfnamefont{A.}~\bibnamefont{Peres}},
  \emph{\bibinfo{title}{Quantum Theory: Concepts and Methods}}
  (\bibinfo{publisher}{Kluwer Academic Publisher}, \bibinfo{year}{1998}).

\bibitem[{\citenamefont{Collins et~al.}(2002)\citenamefont{Collins, Gisin,
  Linden, Massar, and Popescu}}]{Collins02}
\bibinfo{author}{\bibfnamefont{D.}~\bibnamefont{Collins}},
  \bibinfo{author}{\bibfnamefont{N.}~\bibnamefont{Gisin}},
  \bibinfo{author}{\bibfnamefont{N.}~\bibnamefont{Linden}},
  \bibinfo{author}{\bibfnamefont{S.}~\bibnamefont{Massar}}, \bibnamefont{and}
  \bibinfo{author}{\bibfnamefont{S.}~\bibnamefont{Popescu}},
  \bibinfo{journal}{Phys. Rev. Lett.} \textbf{\bibinfo{volume}{88}},
  \bibinfo{pages}{040404} (\bibinfo{year}{2002}).

\bibitem[{\citenamefont{Law et~al.}(2000)\citenamefont{Law, Walmsley, and
  Eberly}}]{Law00}
\bibinfo{author}{\bibfnamefont{C.~K.} \bibnamefont{Law}},
  \bibinfo{author}{\bibfnamefont{I.~A.} \bibnamefont{Walmsley}},
  \bibnamefont{and} \bibinfo{author}{\bibfnamefont{J.~H.}
  \bibnamefont{Eberly}}, \bibinfo{journal}{Phys. Rev. Lett.}
  \textbf{\bibinfo{volume}{84}}, \bibinfo{pages}{5304} (\bibinfo{year}{2000}).

\bibitem[{\citenamefont{Law and Eberly}(2004)}]{Law04}
\bibinfo{author}{\bibfnamefont{C.~K.} \bibnamefont{Law}} \bibnamefont{and}
  \bibinfo{author}{\bibfnamefont{J.~H.} \bibnamefont{Eberly}},
  \bibinfo{journal}{Phys. Rev. Lett.} \textbf{\bibinfo{volume}{92}},
  \bibinfo{pages}{127903} (\bibinfo{year}{2004}).

\bibitem[{Aie()}]{Aiello04_2}
\bibinfo{note}{A. Aiello, J. P. Woerdman, in preparation}.

\bibitem[{\citenamefont{Kliger et~al.}(1990)\citenamefont{Kliger, Lewis, and
  Randall}}]{KligerBook}
\bibinfo{author}{\bibfnamefont{D.~S.} \bibnamefont{Kliger}},
  \bibinfo{author}{\bibfnamefont{J.~W.} \bibnamefont{Lewis}}, \bibnamefont{and}
  \bibinfo{author}{\bibfnamefont{C.~E.} \bibnamefont{Randall}},
  \emph{\bibinfo{title}{Polarized Light in Optics and Spettroscopy}}
  (\bibinfo{publisher}{Academic Press, Inc.}, \bibinfo{year}{1990}).

\bibitem[{\citenamefont{Braunstein et~al.}(1992)\citenamefont{Braunstein, Mann,
  and Revzen}}]{Braunstein92}
\bibinfo{author}{\bibfnamefont{S.~L.} \bibnamefont{Braunstein}},
  \bibinfo{author}{\bibfnamefont{A.}~\bibnamefont{Mann}}, \bibnamefont{and}
  \bibinfo{author}{\bibfnamefont{M.}~\bibnamefont{Revzen}},
  \bibinfo{journal}{Phys. Rev. Lett.} \textbf{\bibinfo{volume}{68}},
  \bibinfo{pages}{3259} (\bibinfo{year}{1992}).

\bibitem[{\citenamefont{Aravind}(1995)}]{Aravind95}
\bibinfo{author}{\bibfnamefont{P.~K.} \bibnamefont{Aravind}},
  \bibinfo{journal}{Phys. Lett. A} \textbf{\bibinfo{volume}{200}},
  \bibinfo{pages}{345} (\bibinfo{year}{1995}).

\bibitem[{\citenamefont{Lee}(1988)}]{LeeBook}
\bibinfo{author}{\bibfnamefont{T.~D.} \bibnamefont{Lee}},
  \emph{\bibinfo{title}{Particle Physics and Introduction to Field Theory}}
  (\bibinfo{publisher}{Harwood Academic Publisher}, \bibinfo{address}{Chur,
  Switzerland}, \bibinfo{year}{1988}), \bibinfo{edition}{revised and updated
  first} ed.

\bibitem[{Lin()}]{Lindner04}
\bibinfo{note}{N. H. Lindner, and D. R. Terno, quant-ph/0403029}.

\bibitem[{\citenamefont{Born and Wolf}(1984)}]{BornWolf}
\bibinfo{author}{\bibfnamefont{M.}~\bibnamefont{Born}} \bibnamefont{and}
  \bibinfo{author}{\bibfnamefont{E.}~\bibnamefont{Wolf}},
  \emph{\bibinfo{title}{Principles of Optics}} (\bibinfo{publisher}{Pergamon
  Press}, \bibinfo{year}{1984}), \bibinfo{edition}{sixth} ed.

\bibitem[{\citenamefont{Mandel and Wolf}(1995)}]{MandelBook}
\bibinfo{author}{\bibfnamefont{L.}~\bibnamefont{Mandel}} \bibnamefont{and}
  \bibinfo{author}{\bibfnamefont{E.}~\bibnamefont{Wolf}},
  \emph{\bibinfo{title}{Optical Coherence and Quantum Optics}}
  (\bibinfo{publisher}{Cambridge University Press}, \bibinfo{year}{1995}),
  \bibinfo{edition}{1st} ed.

\bibitem[{\citenamefont{Fainman and Shamir}(1984)}]{Fainman}
\bibinfo{author}{\bibfnamefont{Y.}~\bibnamefont{Fainman}} \bibnamefont{and}
  \bibinfo{author}{\bibfnamefont{J.}~\bibnamefont{Shamir}},
  \bibinfo{journal}{Appl. Opt.} \textbf{\bibinfo{volume}{23}},
  \bibinfo{pages}{3188} (\bibinfo{year}{1984}).

\bibitem[{\citenamefont{Jauch and Rohrlich}(1955)}]{JauchBook}
\bibinfo{author}{\bibfnamefont{J.~M.} \bibnamefont{Jauch}} \bibnamefont{and}
  \bibinfo{author}{\bibfnamefont{F.}~\bibnamefont{Rohrlich}},
  \emph{\bibinfo{title}{The Theory of Photons and Electrons}}
  (\bibinfo{publisher}{Addison-Wesley Publ. Co.}, \bibinfo{address}{Cambridge,
  MA}, \bibinfo{year}{1955}), \bibinfo{edition}{revised and updated first} ed.

\bibitem[{\citenamefont{Abouraddy et~al.}(2002)\citenamefont{Abouraddy,
  Sergienko, Saleh, and Teich}}]{Abouraddy02}
\bibinfo{author}{\bibfnamefont{A.~F.} \bibnamefont{Abouraddy}},
  \bibinfo{author}{\bibfnamefont{A.~V.} \bibnamefont{Sergienko}},
  \bibinfo{author}{\bibfnamefont{B.~E.~A.} \bibnamefont{Saleh}},
  \bibnamefont{and} \bibinfo{author}{\bibfnamefont{M.~C.} \bibnamefont{Teich}},
  \bibinfo{journal}{Opt. Commun.} \textbf{\bibinfo{volume}{201}},
  \bibinfo{pages}{93} (\bibinfo{year}{2002}).

\bibitem[{\citenamefont{Kwiat et~al.}(1995)\citenamefont{Kwiat, Mattle,
  Weinfurter, Zeilinger, Sergienko, and Shih}}]{Kwiat95}
\bibinfo{author}{\bibfnamefont{P.~G.} \bibnamefont{Kwiat}},
  \bibinfo{author}{\bibfnamefont{K.}~\bibnamefont{Mattle}},
  \bibinfo{author}{\bibfnamefont{H.}~\bibnamefont{Weinfurter}},
  \bibinfo{author}{\bibfnamefont{A.}~\bibnamefont{Zeilinger}},
  \bibinfo{author}{\bibfnamefont{A.~V.} \bibnamefont{Sergienko}},
  \bibnamefont{and} \bibinfo{author}{\bibfnamefont{Y.}~\bibnamefont{Shih}},
  \bibinfo{journal}{Phys. Rev. Lett.} \textbf{\bibinfo{volume}{75}},
  \bibinfo{pages}{4337} (\bibinfo{year}{1995}).

\bibitem[{\citenamefont{Kaku}(1993)}]{KakuBook}
\bibinfo{author}{\bibfnamefont{M.}~\bibnamefont{Kaku}},
  \emph{\bibinfo{title}{Quantum Field Theory. A Modern Introduction}}
  (\bibinfo{publisher}{Oxford University Press}, \bibinfo{address}{Oxford},
  \bibinfo{year}{1993}).

\bibitem[{\citenamefont{Clauser et~al.}(1969)\citenamefont{Clauser, Horne,
  Shimony, and Holt}}]{Clauser69}
\bibinfo{author}{\bibfnamefont{J.~F.} \bibnamefont{Clauser}},
  \bibinfo{author}{\bibfnamefont{M.~A.} \bibnamefont{Horne}},
  \bibinfo{author}{\bibfnamefont{A.}~\bibnamefont{Shimony}}, \bibnamefont{and}
  \bibinfo{author}{\bibfnamefont{R.~A.} \bibnamefont{Holt}},
  \bibinfo{journal}{Phys. Rev. Lett.} \textbf{\bibinfo{volume}{23}},
  \bibinfo{pages}{880} (\bibinfo{year}{1969}).

\bibitem[{\citenamefont{Aiello et~al.}(2003)\citenamefont{Aiello, van Exter,
  and Woerdman}}]{Aiello03}
\bibinfo{author}{\bibfnamefont{A.}~\bibnamefont{Aiello}},
  \bibinfo{author}{\bibfnamefont{M.~P.} \bibnamefont{van Exter}},
  \bibnamefont{and} \bibinfo{author}{\bibfnamefont{J.~P.}
  \bibnamefont{Woerdman}}, \bibinfo{journal}{Phys. Rev. E}
  \textbf{\bibinfo{volume}{68}}, \bibinfo{pages}{046208}
  (\bibinfo{year}{2003}).

\end{thebibliography}
\end{document}